\documentclass[epj]{svjour}
\usepackage{graphics}
\usepackage[T1]{fontenc}
\usepackage[latin1]{inputenc}
\begin{document}
\title{Quadrupole collectivity of neutron-rich Neon isotopes.}
\author{R.R. Rodríguez-Guzmán\inst{1} \and J.L. Egido and L.M. Robledo \inst{2}
}                     
%
\mail{Luis.Robledo@uam.es}
\institute{Institut für Theoretische Physik der Universität Tübingen, Auf der
Morgenstelle 14, D-72076 Tübingen, Germany. \and 
Departamento de Física Teórica C-XI, Universidad Autónoma de Madrid,
28049-Madrid,\\
Spain.}
\date{Received: date / Revised version: date}
%
\abstract{
The angular momentum projected Generator Coordinate Method, with the
quadrupole moment as collective coordinate and the Gogny force (D1S) as
the effective interaction, is used to describe the properties of the
ground state and low-lying excited states of the even-even Neon isotopes
\( ^{20-34} \)Ne, that is, from the stability valley up to the drip-line.
It is found that the ground state of the N=20 nucleus \( ^{30} \)Ne
is deformed but to a lesser extent than the N=20 isotope of the Magnesium.
In the calculations, the isotope \( ^{32} \)Ne is at the drip-line
in good agreement with other theoretical predictions. On the other
hand, rather good agreement with experimental data for many observables
is obtained. 
} 
\maketitle

\section{Introduction}

The properties of the ground and lowest-lying excited states of nuclei close to
the stability valley are determined to a great extent by the underlying mean
field in which all the nucleons move. The nuclear mean field, obtained through
the mean field approximation to the nuclear many body problem, provides the
important concepts of magic numbers (or shell closures) and that of spontaneous
symmetry breaking. For nuclei with proton and/or neutron numbers close to the
magic ones symmetry conserving (i.e. non superconducting and spherical) ground
states are expected. On the other hand, nuclei away from the magic
configurations are expected to show strong symmetry breaking at the mean field
level leading to deformed (and superconducting) ground states
which are the heads of bands generated kinematically by restoring the broken
symmetries -like the well known example of the rotational bands.

Almost thirty years ago, experimental evidences were found in the neutron-rich
light nuclei around \( ^{31} \)Na \cite{Thibault.75,Touchard.82} pointing to 
the breaking of the N=20 magic number. The first attempt to understand the
phenomenon from a theoretical point of view \cite{Campi.75} suggested that
correlations beyond mean field coming from the restoration of the rotational
symmetry could be strong enough as to overcome the mean field shell  effects
and lead to deformed ground states for closed shell nuclei like \( ^{31} \)Na
and \( ^{32} \)Mg. Further theoretical studies using the  Shell Model (SM)
approach were carried out \cite{Poves.87,Warburton.90} and it was suggested
that an intruder configuration consisting of a two particle two hole neutron
excitation from the sd shell into the f\( _{7/2} \) one was  the responsible
for deformation in the ground state of \( ^{32} \)Mg.

The recent availability of Radioactive Ion Beam facilities in several
laboratories like Ganil, GSI, MSU and Riken to cite a few and the development
of very efficient mass separators and solid state detectors has made possible
to measure up many properties  concerning the ground state and the lowest lying
excited states of many exotic, neutron-rich light nuclei. In particular, the
exploration of both the \( N=20 \) and \( N=28 \) shell closures far from
stability has proven to be a rich source of new phenomena. Among the variety of
available experimental data, the most convincing evidence for a deformed ground
state in the region around \( N=20 \) is found in the \( ^{32} \)Mg nucleus
where both the excitation energies of the lowest lying \( 2^{+}
\)\cite{32Mg.E2+} and \( 4^{+} \) \cite{32Mg.E4+} states and the \(
B(E2,0^{+}\rightarrow 2^{+}) \) transition probability~\cite{32Mg.BE2} have
been measured. The low excitation energy of the \( 2^{+} \) state, the high
value of the \( B(E2) \) transition probability and also the ratio \(
E(4_{1}^{+})/E(2_{1}^{+})=2.6 \) are fairly compatible with the expectations
for a rotational band. Additional evidence comes from the neighboring isotope
\( ^{34} \)Mg where the \( E(2^{+}_{1}) \) is only 0.66 MeV, the \(
B(E2,0^{+}\rightarrow 2^{+}) \) is 631 \( e^{2}\textrm{fm}^{2} \) and the \(
E(4_{1}^{+})/E(2_{1}^{+}) \) ratio is 3.2, that is, the \( 0^{+}_{1} \), \(
2^{+}_{1} \) and \( 4^{+}_{1} \) satisfy all the requirements to belong to a
strongly deformed rotational band \cite{Yoneda.01,Iwa.01}.

From a theoretical point of view, the ground state of \( ^{32} \)Mg is
spherical at the mean field level. However, when the zero point rotational
energy correction (ZPRE) stemming from the restoration of the rotational
symmetry is considered, the energy landscape as a function of the quadrupole
moment changes dramatically and \( ^{32} \)Mg becomes deformed
\cite{Campi.75,Berger.93,Rein.99,Heenen.99,Rayner.2000}.
A more careful analysis of the energy landscape including the ZPRE correction
reveals that, in fact, there are two coexistent configurations (prolate and
oblate) with similar energies indicating thereby that configuration mixing of
states with different quadrupole intrinsic deformation has to be considered.
Therefore, an Angular Momentum Projected Generator Coordinate Method (AMP-GCM)
calculation with the quadrupole moment as collective coordinate is called for.
We have applied this method in Ref. \cite{Rayner.2000.a} to the study of the
nuclei \( ^{30-34} \)Mg with the Gogny force. We have obtained prolate ground
states for \( ^{32-34} \)Mg indicating that the \( N=20 \) shell closure is not
preserved for the Mg isotopes. Moreover, a good agreement with the experimental
data for the \( 2^{+} \) excitation energies and \( B(E2) \) transition
probabilities was obtained. The method has been used to study quadrupole
collectivity of the Si isotopes \cite{Rayner.2001}, the N=28 isotopes
\cite{N=28} as well as the superdeformed band in \( ^{32} \)S \cite{ref_32S}
with quite a good success.

The purpose of this paper is to extend the AMP-GCM calculations to the study of
the Neon (two proton less than Magnesium) isotopes from A=20 up to  A=34.
Contrary to the Magnesium isotopic chain, the Neon neutron drip line (N=22) is
very close to the neutron magic number N=20 and therefore the study of the Neon
isotopes is a good testing ground to examine both the systematic of deformation
and the possible erosion of the N=20 spherical shell closure very close to the
neutron drip line. Another interesting point is the study of the possible magicity
of N=16 suggested in recent analyses  \cite{Ozawa:2000}.

The paper is organized as follows: In section 2 a brief overview of the
theoretical framework is presented. In section 3.1 the mean field results are
discussed. In the next section the effect of angular momentum projection on the
mean field observables is described. Finally, in section 3.3 the results of the
configuration mixing calculations are presented and compared to the experimental
data and other theoretical approaches. We end up with the conclusions in section
4.

\section{Theoretical framework}

\label{THEORY}
As mentioned in the introduction ours is a mean field
based procedure where the underlying mean field is determined first
and then additional correlations beyond mean field are included. Those
additional correlations are handled in the framework of the angular
momentum projected Generator Coordinate Method (AMP-GCM) with the
mass quadrupole moment as generating coordinate. As we restrict ourselves
to axially symmetric configurations, we use the following ansatz for
the \( K=0 \) AMP-GCM wave functions 
\begin{equation}
\left| \Phi ^{I}_{\sigma }\right\rangle =\int dq_{20}f^{I}_{\sigma }(q_{20})\hat{P}^{I}_{00}\left| \varphi (q_{20})\right\rangle .
\end{equation}
For each angular momentum \( I \) the different AMP-GCM states (labelled
by \( \sigma  \)) are linear combinations of the set of angular momentum
projected intrinsic wave functions \( \left| \varphi (q_{20})\right\rangle  \)
generated by solving the Hartree-Fock-Bogoliubov
(HFB) equation constrained to yield the desired
mass quadrupole moment 
\( q_{20}=\left\langle \varphi (q_{20})\right| z^{2}-1/2(x^{2}+y^{2})\left| \varphi (q_{20})\right\rangle  \). 

The intrinsic wave functions are restricted to be axially symmetric (i.e. \(
K=0 \)) and are obtained by solving the HFB equation with  the Gogny
interaction ~\cite{Dech_Gogny.80}  (D1S parameterization~\cite{Berger.84_D1S}).
The HFB equation is discretized by expanding the quasiparticle operators
associated to the intrinsic wave functions \( \left| \varphi
(q_{20})\right\rangle  \) in a Harmonic Oscillator (HO) basis containing eleven
major shells and with equal oscillator lengths (in order to make the basis
closed under rotations \cite{Robledo.94}). As we are dealing with quite light
systems we have to consider the center of mass problem. This is handled by
subtracting the center of mass kinetic energy both in the calculation of the
energy and in the HFB variational procedure. Finally, concerning the Coulomb
interaction, we have only taken into account its contribution to the direct
field in the variational procedure. The exchange Coulomb energy (computed in
the Slater approximation) is added, in a perturbative fashion, at the end of
the calculation and the contribution of the Coulomb interaction to the pairing
energy is completely disregarded.

In order to obtain the angular momentum projected wave functions we
use the standard angular momentum projector operator restricted to
\( K=0 \) states \cite{Hara.95} 
\begin{equation}
\hat{P}^{I}_{00}=\frac{(2I+1)}{8\pi ^{2}}\int d\Omega d^{I}_{00}(\beta)
e^{-i\alpha \hat{J}_{z}}e^{-i\beta \hat{J}_{y}}e^{-i\gamma \hat{J}_{z}}.
\end{equation}
Finally, the ``collective amplitudes'' \( f^{I}_{\sigma }(q_{20}) \)
as well as the energies of the AMP-GCM states \( \left| \Phi ^{I}_{\sigma }\right\rangle  \)
are obtained through the solution of the Hill-Wheeler (HW) equation
\begin{equation}
\int dq'_{20}{\mathcal{H}}^{I}(q_{20},q'_{20})f^{I}_{\sigma }(q'_{20})=
E^{I}_{\sigma }
\int dq'_{20}\mathcal{N}^{I}(q_{20},q'_{20})f^{I}_{\sigma }(q'_{20})
\end{equation}
which is given in terms of the projected norm 
\begin{equation}
{\mathcal{N}}^{I}(q_{20},q'_{20})=
\left\langle \varphi (q_{20})\right| \hat{P}^{I}_{00}\left| \varphi (q'_{20})\right\rangle 
\end{equation}
 and the projected hamiltonian kernel 
 \begin{equation} \label{hamover}
{\mathcal{H}}^{I}(q_{20},q'_{20})=
\left\langle \varphi (q_{20})\right| \hat{H}\hat{P}^{I}_{00}\left| \varphi (q'_{20})\right\rangle .
\end{equation}
As the generating states \( \hat{P}^{I}_{00}\left| \varphi
(q_{20})\right\rangle  \) are not orthogonal, the ``collective amplitudes'' \(
f^{I}_{\sigma }(q_{20}) \) cannot be interpreted as probability amplitudes.
Instead, one usually introduce \cite{Ring_Suck.80} the ``collective" wave
functions 
\begin{equation}
g^{I}_{\sigma }(q_{20})=\int dq'_{20}f^{I}_{\sigma }(q'_{20})\left( \mathcal{N}^{I}(q_{20},q'_{20})^{*}\right) ^{1/2}
\end{equation}
 which are orthonormal \( \int dq_{20}{g^{I}_{\sigma}}^*(q_{20})g^{I'}_{\sigma '}(q_{20})=\delta _{I,I'}\delta _{\sigma ,\sigma '} \)
and therefore their module squared has the meaning of a probability.

In order to readjust the particle number on the average for the projected wave
functions the Hamiltonian in Eq. (\ref{hamover}) has been replaced by 
$\hat{H}'=\hat{H}-\lambda_n (\hat{N}-N_0) -\lambda_p (\hat{Z}-Z_0)$ where
$\lambda_n$ and $\lambda_p$ are chemical potential parameters (see Ref 
\cite{Rayner.2002}, and references therein).

\begin{figure*}
{\centering \resizebox*{0.8\textwidth}{!}{\includegraphics{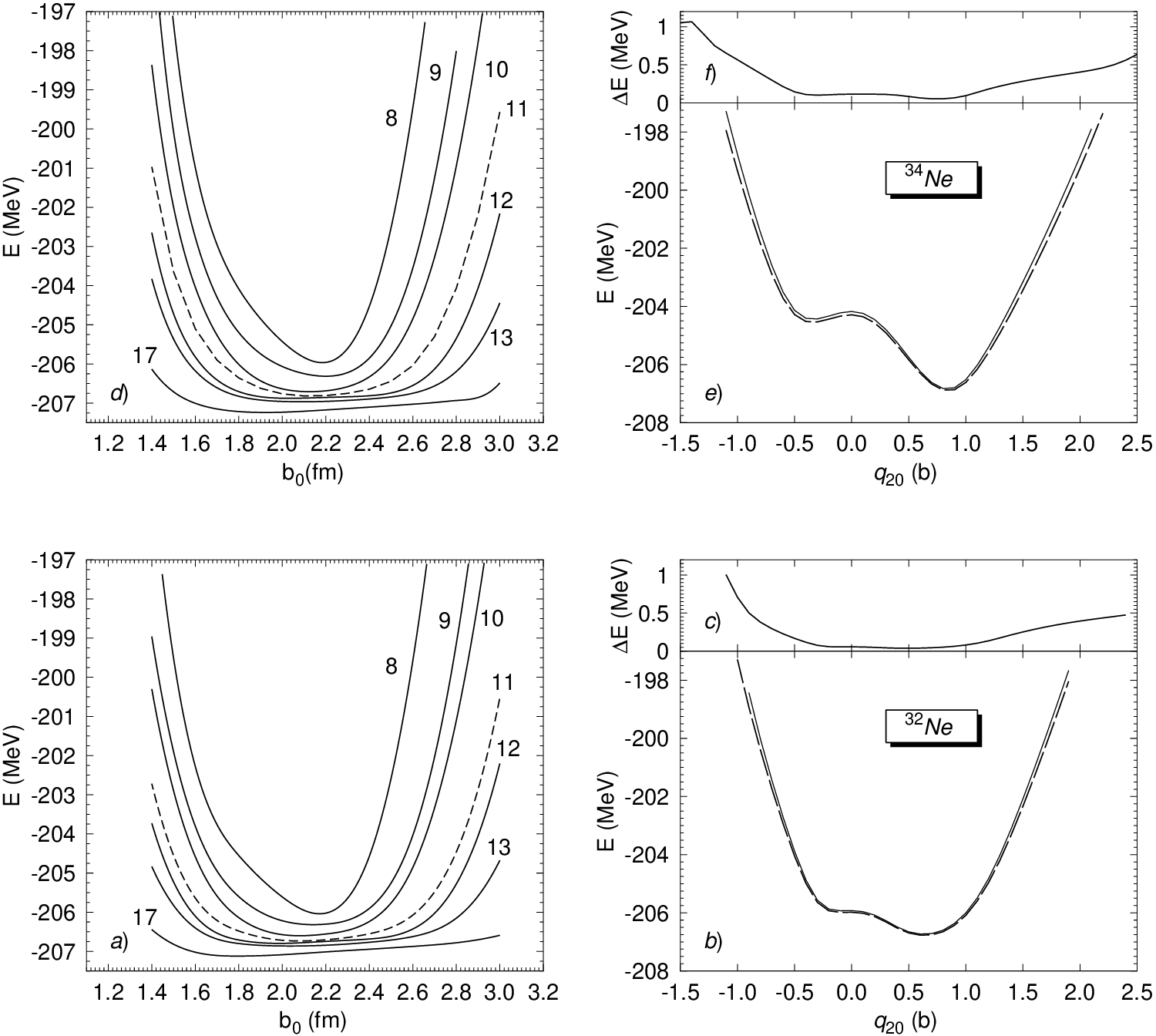}} \par}

\caption{In panels a) and d) the ground state HFB energies of \protect\( ^{32,34}\protect \)Ne
are plotted as functions of the HO length \protect\( b_{0}\protect \)
(\protect\( b_{\perp }=b_{z}=b_{0})\protect \) for the \protect\( N_{shell}=8,9,10,11,12,13\protect \)
and 17 bases. The curve corresponding to the basis used in the present
work (\protect\( N_{shell}=11\protect \)) is plotted as a dashed
line. In panels b) and e) the HFB energies of \protect\( ^{32,34}\protect \)Ne
computed with \protect\( b_{0}=2.1\protect \)fm and \protect\( N_{shell}=11\protect \)(full
line) and \protect\( N_{shell}=12\protect \) (dashed line) are plotted
as functions of the quadrupole moment. In panels c) and f), the energy
differences between \protect\( N_{shell}=11\protect \) and \protect\( N_{shell}=12\protect \)
calculations are plotted as functions of the quadrupole moment.}
\label{Fig_01}
\end{figure*}

The \( B(E2) \) transition probabilities are computed using the AMP-GCM
wave functions
\begin{eqnarray}
&& B(E2,I_{i}\rightarrow I_{f})=\frac{e^{2}}{2I_{i}+1} \times \\ \nonumber 
&&\left| \int dq_{20}dq'_{20}f^{I_{f}*}_{\sigma _{f}}(q'_{20})\langle I_{f}q'_{20}\mid \mid \hat{Q}_{2}\mid \mid I_{i}q_{20}\rangle f^{I_{i}}_{\sigma _{i}}(q_{20})\right| ^{2}
\end{eqnarray}
 with
\begin{eqnarray*}
&&\frac{\langle I_{f}q'_{20}\mid \mid \hat{Q}_{2}\mid \mid I_{i}q_{20}\rangle }{(2I_{i}+1)(2I_{f}+1)}=\sum _{\mu }\left( \begin{array}{ccc}
I_{i} & 2 & I_{f}\\
-\mu  & \mu  & 0
\end{array}\right) \times \\ 
&&\int _{0}^{\frac{\pi }{2}}d\beta \sin \beta d_{-\mu ,0}^{I_{i}}(\beta )\langle \varphi (q'_{20})\mid \hat{Q}_{2\mu }e^{-i\beta \hat{J}_{y}}\mid \varphi (q_{20})\rangle 
\end{eqnarray*}
where the indices i and f stand for the initial and final states
and \( \hat{Q}_{2\mu } \) are the charge quadrupole moment operators.
As we are using the full configuration space no effective charges
are needed. Further details on the computational procedure can be
found in Ref \cite{Rayner.2002}.

\section{Discussion of the results}

\subsection{Mean field approximation.}

\label{MF_RESULTS} 
Before discussing the mean field results, the 
convergence of our calculations with the size
of the Harmonic Oscillator (HO) basis used to discretize the HFB equation
has to be tested as we are dealing with near drip-line nuclei like \( ^{32,34} \)Ne.
First, one should note that, due to the proximity of the drip-line, the full
HFB approximation must be used \cite{Doba.84,Doba.96}. It is also evident that
absolute convergence for the binding energies can only be attained for HO basis
with an infinity number of shells (\( N_{shell} \)). However, it is expected
that other physical observables like excitation energies, transition
probabilities, etc can be accurately described with a finite number of shells.

In order to determine how many shells are needed for an accurate description of
the physical observables we have first studied the behavior of the HFB energy
as a function of the oscillator length parameter \( b_{0} \) (\( b_{\perp
}=b_{z}=b_{0}) \) and the number of shells for the nuclei $^{32}$Ne and
$^{34}$Ne. The results are plotted in panels a) and d) of Fig. \ref{Fig_01}. As
expected the curves become  more and more flat for increasing values of \(
N_{shell} \) and already the \( N_{shell}=17 \) basis can be considered as a
good approximation for an infinite basis in both isotopes as the dependence of
the energies on the oscillator length is very weak for a wide range of \( b_{0}
\) values. However, \( N_{shell}=17 \) makes the HO basis too big for our
purposes as we have to  evaluate the AMP-GCM hamiltonian kernel which is a
quantity requiring two orders of magnitude more computing time than the whole
HFB calculation. We have found that \( N_{shell}=11 \) is a good compromise
between accuracy and computing time as the energy still shows a weak dependency on the
oscillator length around the minima located at \( b_{0}=2.1 \) fm for both \(
^{32} \)Ne (panel a) and \( ^{34} \)Ne (panel d).  
As an example of the adequacy of the \( N_{shell}=11 \) calculations let us
consider the two neutron separation energy of \( ^{34} \)Ne.
The \(
N_{shell}=17 \) ground state energies for \( ^{32} \)Ne and \( ^{34} \)Ne are
403 keV and 420 keV lower than the \( N_{shell}=11 \) energies implying that
the two neutron separation energy of \( ^{34} \)Ne computed with \(
N_{shell}=11 \) differs by 17 keV from the one computed with \( N_{shell}=17
\).

In order to study the suitability of the \( N_{shell}=11 \) basis for other
nuclear properties we have performed calculations with \( N_{shell}=12 \). The
mean field results for both calculations  are compared on the right hand side
panels of Fig. \ref{Fig_01}.  In panels b) and e) the energy landscapes of \(
^{32} \)Ne and \( ^{34} \)Ne are shown as a function of the quadrupole moment
for the \( N_{shell}=11 \) and \( N_{shell}=12 \) calculations. In the upper
panels c) and f) we have represented the energy differences between both
calculations. The conclusion is that in the region  \( -0.7\textrm{b}\leq
q_{20}\leq 1.2\textrm{b} \) the shape of the energy landscapes do not change
much when the basis is increased from \( N_{shell}=11 \) to \( N_{shell}=12 \).
As we will see later on this is the range of \( q_{20} \) values
where the collective dynamics is concentrated and therefore it is not expected
to find significant differences between the \( N_{shell}=11 \) and \(
N_{shell}=12 \) results. 

\begin{figure}
{\centering \resizebox*{0.45\textwidth}{!}{\includegraphics{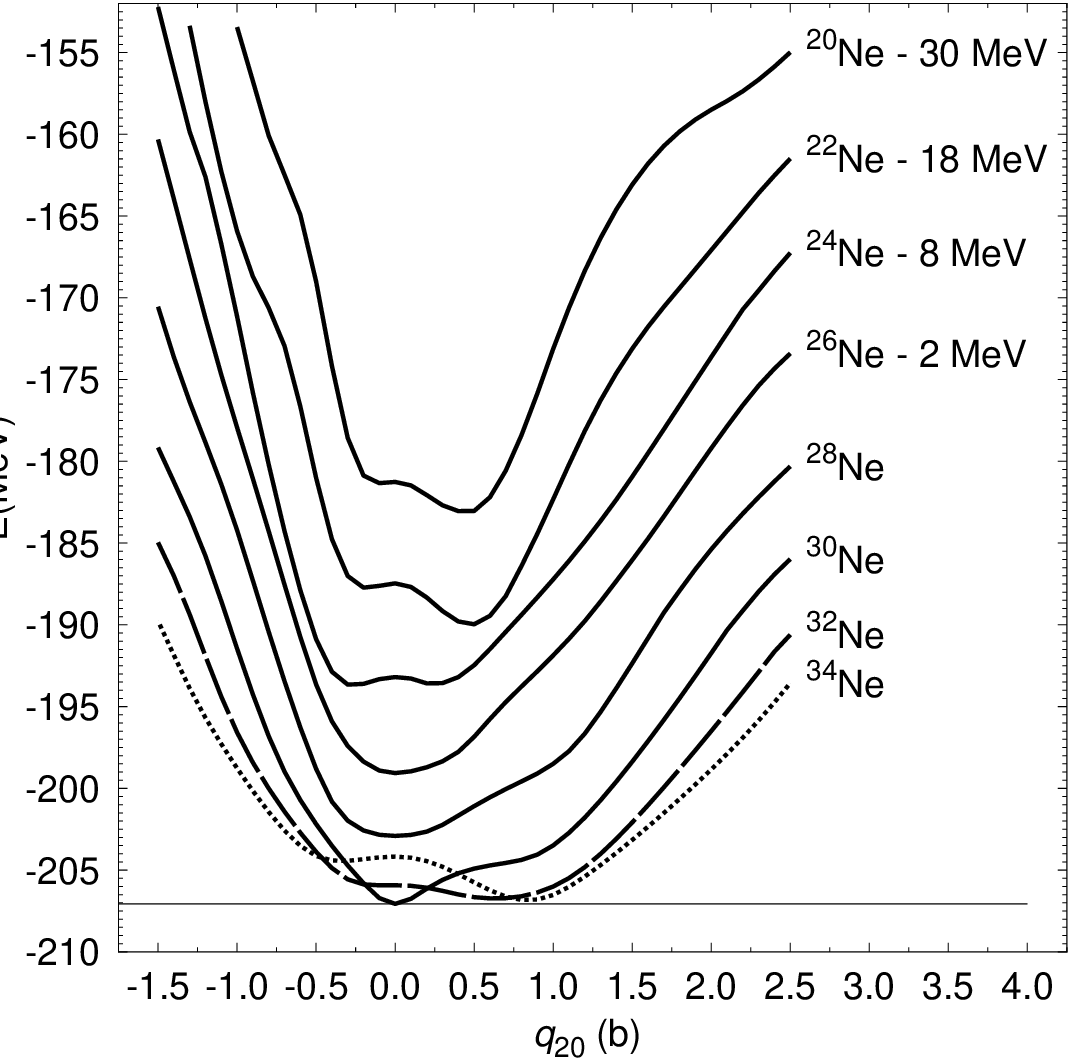}} \par}

\caption{Mean field potential energy surfaces for the considered
Neon isotopes, plotted as a function of the axially symmetric quadrupole
moment. The curves have been shifted to show them in a single plot.
The corresponding energy shifts are given in the plot.}\label{Fig:MFPES}
\end{figure}

\begin{figure}
\resizebox*{0.45\textwidth}{!}{\includegraphics{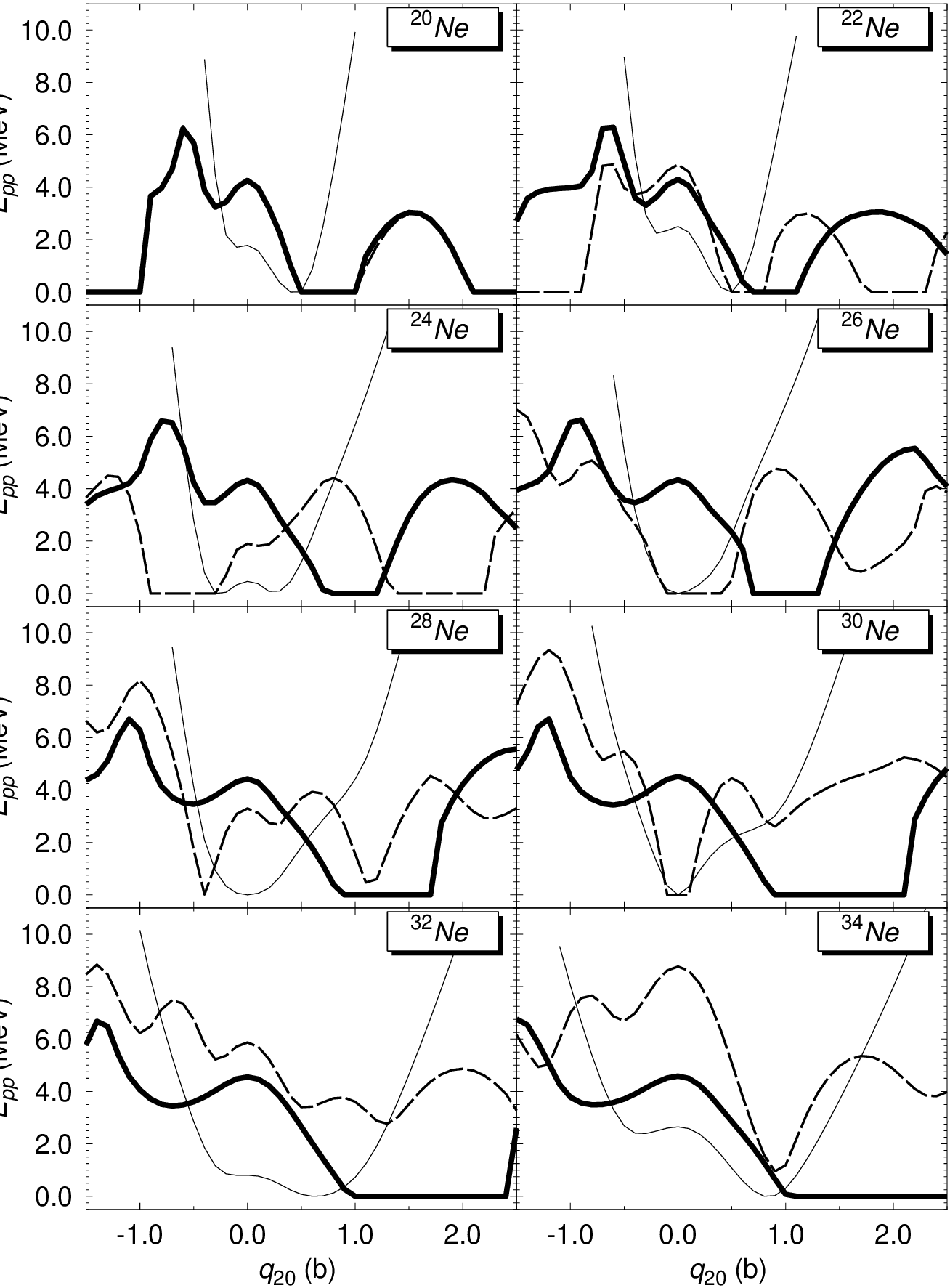}} 

\caption{In each panel, proton (thick full lines) and neutron (thick dashed
lines) particle-particle energies \protect\( -E_{pp}\protect \) are
depicted as a function of the quadrupole moment. Also the energy differences
\protect\( E_{\textrm{HFB}}(q_{20})-E_{\textrm{g}.\textrm{s}.}\protect \),
where \protect\( E_{\textrm{g}.\textrm{s}.}\protect \) corresponds
to the ground state HFB configuration, are plotted as thin full lines.}

\label{Fig:Pairing}
\end{figure}

In Fig. \ref{Fig:MFPES} the mean field potential energy surfaces
(MFPES) are plotted as a function of the axially symmetric quadrupole
moment \( q_{20} \) for the even-even Neon isotopes \( ^{20-34} \)Ne.
The MFPES shown do not include the Coulomb exchange energy and they
have been shifted to accommodate them in a single plot (see figure caption). 

Both \( ^{20} \)Ne and \( ^{22} \)Ne are prolate deformed in their
ground states. In \( ^{20} \)Ne the prolate ground state corresponds
to \( q_{20}=0.4 \)b \( (\beta _{2}=0.37) \) and an oblate local
minimum also appears at \( q_{20}=-0.1 \)b \( (\beta _{2}=-0.09) \)
with an excitation energy of \( 1.71 \) MeV. In the case of \( ^{22} \)Ne
the ground state corresponds to \( q_{20}=0.5 \)b \( (\beta _{2}=0.40) \)
and another local minimum is found at \( q_{20}=-0.2 \)b \( (\beta _{2}=-0.17) \)
with an excitation energy of \( 2.24 \) MeV. The nucleus \( ^{24} \)Ne
is a clear example, in the considered isotopic chain, of very strong
shape coexistence since, while the oblate ground state is locate at
\( q_{20}=-0.3 \)b \( (\beta _{2}=-0.22) \), a prolate isomeric
state is also found at \( q_{20}=0.2 \)b \( (\beta _{2}=0.15) \)
with an excitation energy with respect to the oblate ground state
of 77 keV. On the other hand, the nuclei \( ^{26-30} \)Ne show spherical
ground states indicating that the N=20 shell closure is preserved
at the mean field level. The MFPES of both \( ^{26,28} \)Ne are particularly
flat around their spherical ground states. In the nucleus \( ^{30} \)Ne
we obtain a prolate shoulder at \( q_{20}=0.8 \)b \( (\beta _{2}=0.37) \)
at an excitation energy of 2.70 MeV with respect to the spherical
ground state. This prolate shoulder is around 1
MeV higher than the one found in similar HFB calculations in\( ^{32} \)Mg
(see for example \cite{Rayner.2000.a}). In the drip-line systems \( ^{32} \)Ne
and \( ^{34} \)Ne, prolate deformed ground states are found. The
ground states have \( q_{20}=0.6 \)b \( (\beta _{2}=0.26) \) and
\( q_{20}=0.8 \)b \( (\beta _{2}=0.31) \), respectively. In addition,
an oblate isomeric state is found in \( ^{34} \)Ne at \( q_{20}=-0.3 \)b
\( (\beta _{2}=-0.12) \) with an excitation energy of 2.39 MeV with
respect to the prolate ground state.

Another interesting point concerns the stability of the Ne isotopes against
neutron emission. From the absolute minimum of the MFPES' depicted
in Fig. \ref{Fig:MFPES} (the one of $^{30}$Ne
is marked with a horizontal line) it  can be deduced that \( ^{32} \)Ne is not
stable against two neutron emission. This is in clear contradiction with the
experimental results as they indicate that \( ^{32} \)Ne is indeed stable
against two neutron emission \cite{GANIL_1,GANIL_2,SAKU_1}. On the other hand,
the  nucleus $^{34}$Ne is slightly stable against two neutron emission, but it
is not against four neutron emission.

In Fig. \ref{Fig:Pairing} the proton and neutron particle-particle correlation
energies \( -E_{pp}=\frac{1}{2}Tr\left( \Delta \kappa ^{*}\right)  \), which
are commonly used to discuss pairing correlations (see for example Refs. 
\cite{Dech_Gogny.80,Pairing}), are plotted as a function of the quadrupole
deformation for all the isotopes considered. The evolution of the
particle-particle correlation energies is well correlated with the structures
found in the MFPES. Non-zero proton pairing correlations are found in all the
spherical or oblate minima. In addition, sizeable neutron pairing correlations
are found in \( ^{20,22} \)Ne and \( ^{32,34} \)Ne for the spherical and the
oblate minima. Vanishing proton pairing correlations are found in the prolate
side in a window starting at \( 0.5 \)b and ending in 1.0 b in $^{20}$Ne. For
the other isotopes the starting and ending points increase with the neutron
number. Neutron pairing correlations vanish in the ground states of both \(
^{26} \)Ne and \( ^{30} \)Ne. In the later case this is a consequence of the
N=20 shell closure found at the mean field level.

It should be stressed here that the unphysical collapse of pairing
correlations found in Fig. \ref{Fig:Pairing} is an indication that
one should also consider in these isotopes  dynamical pairing correlations 
and their coupling to the quadrupole degree of freedom in
the scope of a formalism beyond the mean field in order to treat them
in an equal footing. However, this multidimensional configuration
mixing calculation is a cumbersome task with the present-day computational
facilities.

\subsection{Correlations beyond the mean field: Angular momentum projection.}

\begin{figure}
{\centering\resizebox*{0.45\textwidth}{!}{\includegraphics{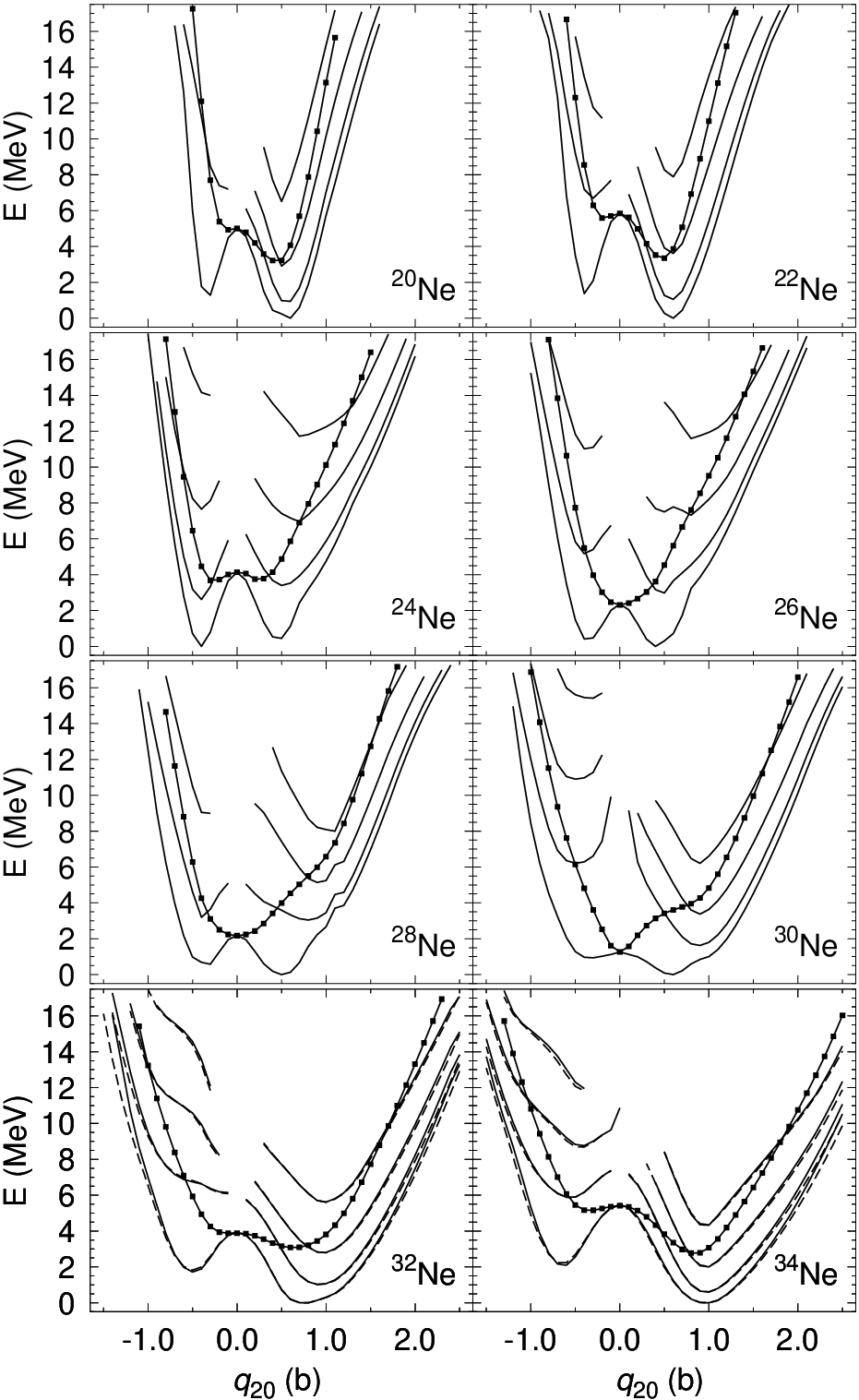}} \par}

\caption{Angular momentum projected potential energy surfaces (full
lines) for the nuclei \protect\( ^{20-34}\protect \)Ne and for the
spin values \protect\( I^{\pi }=0^{+},2^{+},4^{+}\protect \)and \protect\( 6^{+}\protect \),
plotted as a function of the axially symmetric quadrupole moment \protect\( q_{20}\protect \).
The mean field potential energy surfaces are also plotted as lines
with boxes. In each nucleus, the energies are referred to the energy
of the \protect\( I^{\pi }=0^{+}\protect \) ground state. For the
nuclei \protect\( ^{32-34}\protect \)Ne we have also included (dashed
lines) the projected results corresponding to the calculation with
\protect\( N_{shell}=12\protect \).}\label{Fig:AMPPES}
\end{figure}

Before considering the full AMP-GCM it is instructive to look into
the angular momentum projected energy surfaces (AMPPES) defined as
\begin{equation}
E^{I}(q_{20})=\frac{{\mathcal{H}}^{I}(q_{20},q_{20})}
                   {{\mathcal{N}}^{I}(q_{20},q_{20})}
\end{equation}
and shown in Fig. \ref{Fig:AMPPES} for the nuclei \( ^{20-34} \)Ne
and \( I=0 \), 2, 4, and 6. The corresponding mean field energy landscapes
(dashed curves with boxes) are also included for comparison. 
For details on the
missing points in the \( I=2 \), 4 and 6 curves refer to \cite{Rayner.2002}.
The most remarkable fact about Fig. \ref{Fig:AMPPES} is
how strongly the restoration of the rotational symmetry
modifies the mean field picture of the \( I=0 \) configurations.
Contrary to the mean field case, two minima, one prolate and the other
oblate, are obtained for all the Neon isotopes for \( I^{\pi }=0^{+} \).
The prolate minimum is, with the exception of \( ^{24} \)Ne, the
absolute minimum in all the isotopes considered. For increasing spin
values either the energy difference between the prolate and oblate
minima increases or the oblate minimum is washed out.

\begin{figure}
{\centering \resizebox*{0.5\textwidth}{!}{\includegraphics{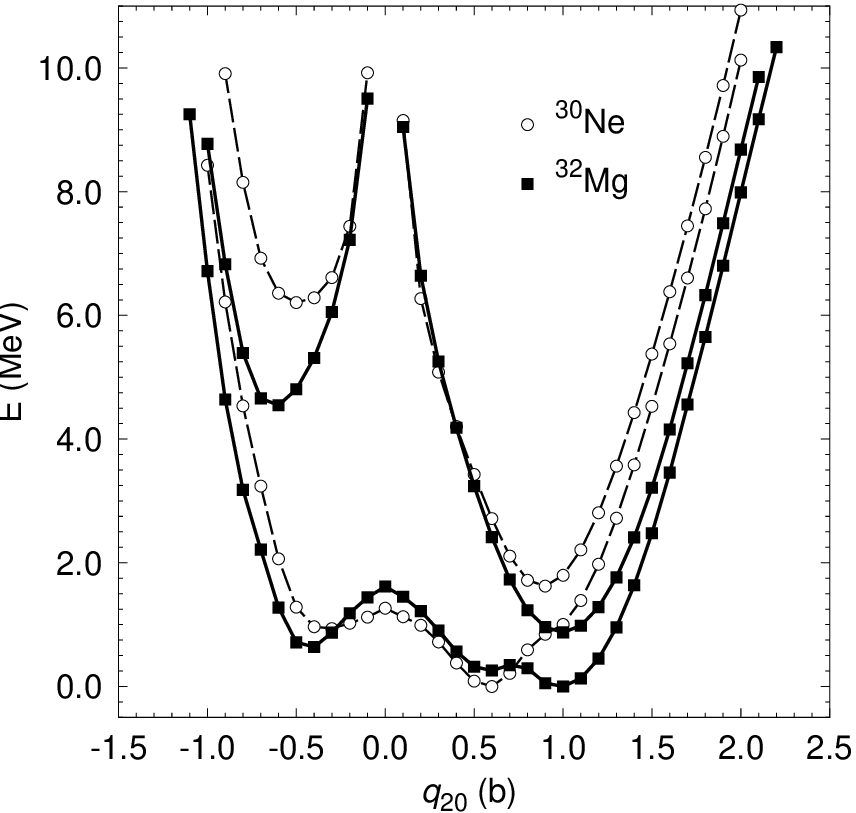}} \par}

\caption{A comparison of the angular momentum projected energies
of the \protect\( ^{30}\protect \)Ne and \protect\( ^{32}\protect \)Mg
nuclei as functions of the quadrupole moment for \protect\( I^{\pi }=0^{+}\protect \)and
\protect\( 2^{+}\protect \)(referred to the \protect\( I^{\pi }=0^{+}\protect \)absolute
minimum in each nucleus).}\label{Fig:30Neand32Mg}
\end{figure}

The nucleus \( ^{24} \)Ne, with its \( I^{\pi }=0^{+} \) oblate
ground state is the exception in all the Neon isotopes studied. The
orbital responsible for such an oblate minimum is the neutron 
\( d_{\frac{5}{2}} \)
orbital which becomes fully occupied in this nucleus and favors oblate
deformations. The absolute minimum remains oblate for \( I^{\pi }=2^{+} \) but
already at \( I^{\pi }=4^{+} \) it becomes prolate deformed.

In addition, shape coexistence is expected in the nuclei \( ^{26} \)Ne,
\( ^{28} \)Ne and \( ^{30} \)Ne as their \( I^{\pi }=0^{+} \) prolate 
and oblate minima are very close in energy (414,
577 and 936 keV, respectively). These minima are separated by barriers
which are 2.4, 2.2 and 1.3 MeV high, respectively. 

Finally, it is also worth to compare the intrinsic quadrupole deformation
of the ground state of \( ^{30} \)Ne with the one of \( ^{32} \)Mg
\cite{Rayner.2000}. In Fig. \ref{Fig:30Neand32Mg}  we have plotted the 
\( I^{\pi }=0^{+} \) and \( 2^{+} \) AMPPES for both \( ^{30} \)Ne 
and \( ^{32} \)Mg. From this plot we conclude that the absolute minimum
of the \( I^{\pi }=0^{+} \) AMPPES in \( ^{30} \)Ne has half the
deformation of the corresponding minimum in \( ^{32} \)Mg. On the
other hand, the \( I^{\pi }=2^{+} \) deformations are practically
identical. From this results one may expect, as is the case, very
similar spectroscopic quadrupole moments for the first \( 2^{+} \)
states in \( ^{30} \)Ne and \( ^{32} \)Mg and a reduction of the
\( B(E2) \) transition probability in \( ^{30} \)Ne as compared to the one of
\( ^{32} \)Mg.

\subsection{Correlations beyond the mean field: Angular momentum projection and
configuration mixing.}

\begin{figure}
{\centering \resizebox*{8cm}{!}{\includegraphics{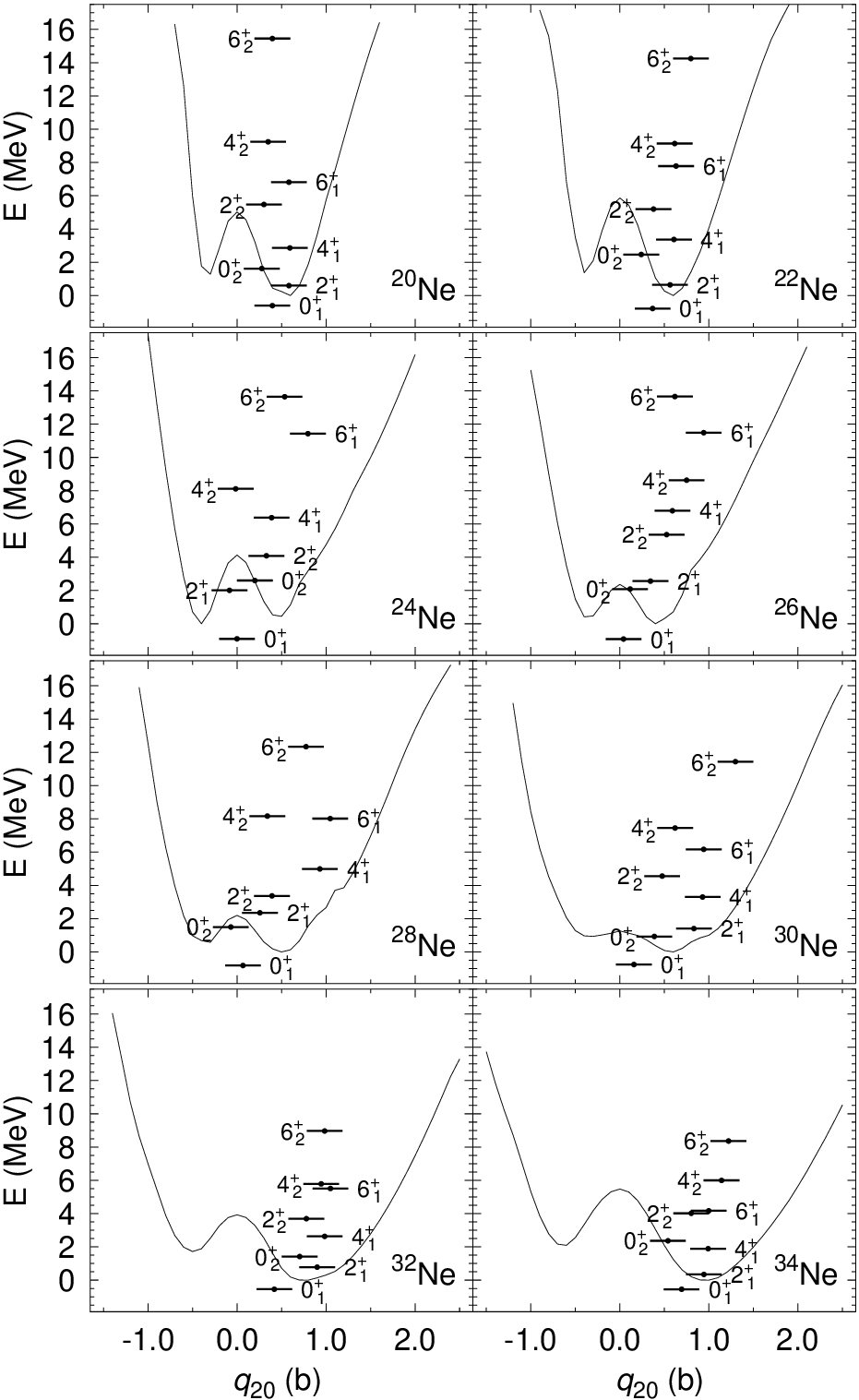}} \par}

\caption{Band diagrams for the nuclei \protect\( ^{20-34}\protect \)Ne obtained
by solving the Hill-Wheeler equation of the AMPGCM. The 
\protect\(I=0\protect \) AMPPES are also plotted to guide the eye. 
See text for further details.\label{Fig:BandDiagram}}
\end{figure}

The AMPPES  of the previous subsection show for some nuclei 
and/or some spin values the phenomenon of shape coexistence and therefore
configuration mixing has to be considered
in order to gain a better understanding of the structure of these states. 
We have considered
configuration mixing in the framework of the Angular Momentum Projected 
Generator Coordinate Method (AMP-GCM)  described in section
\ref{THEORY}. The intrinsic axial quadrupole moment \( q_{20} \) with values
in the range 
\( -1.5\textrm{b}\leq q_{20}\leq 2.5\textrm{b} \) and  with
a mesh size $\Delta q_{20}$ of $10\textrm{fm}^{2}$ has been chosen
as generating coordinate.

\begin{figure*}[bt]
{\centering
\resizebox*{0.95\textwidth}{!}{\includegraphics{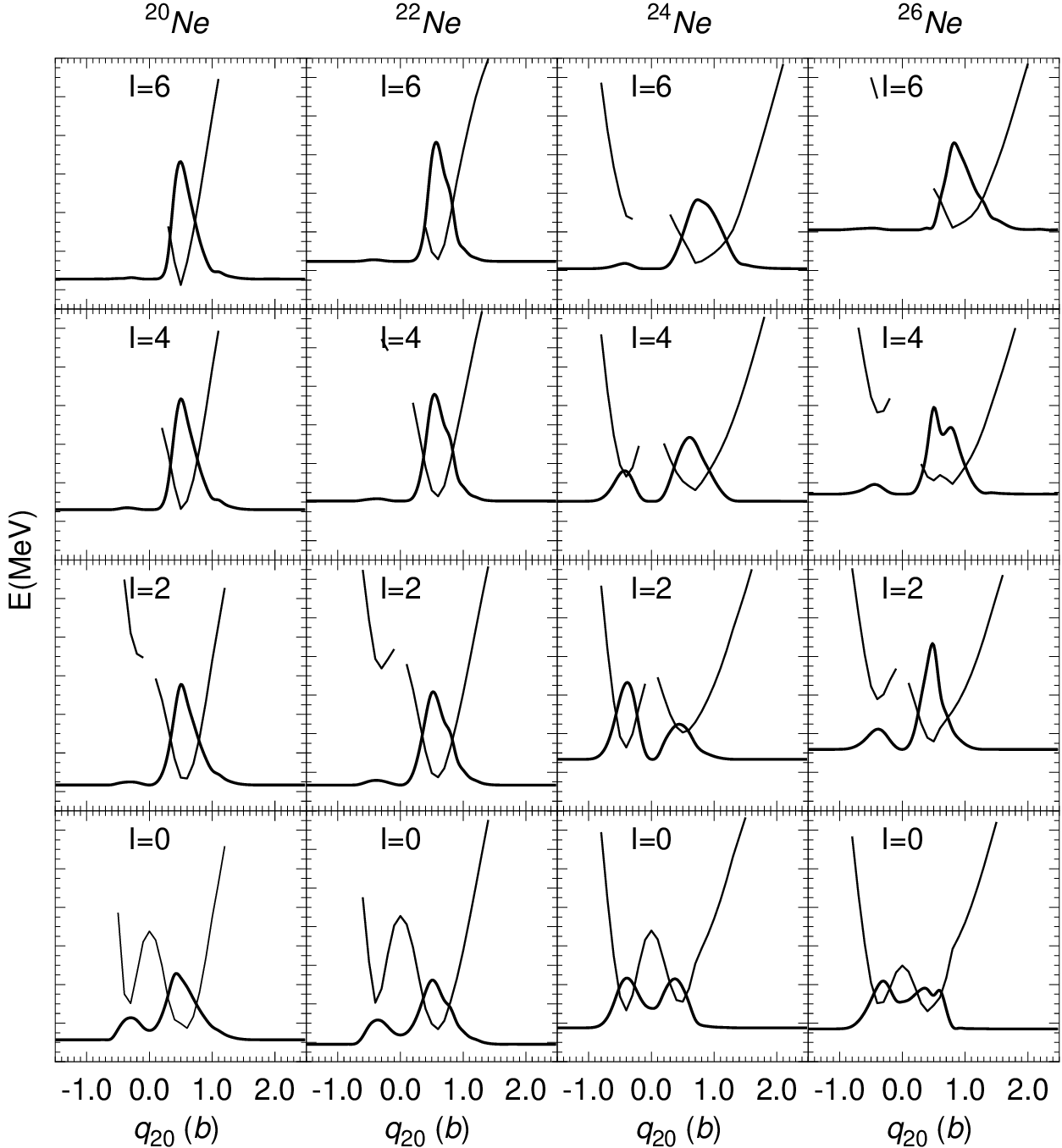}\includegraphics{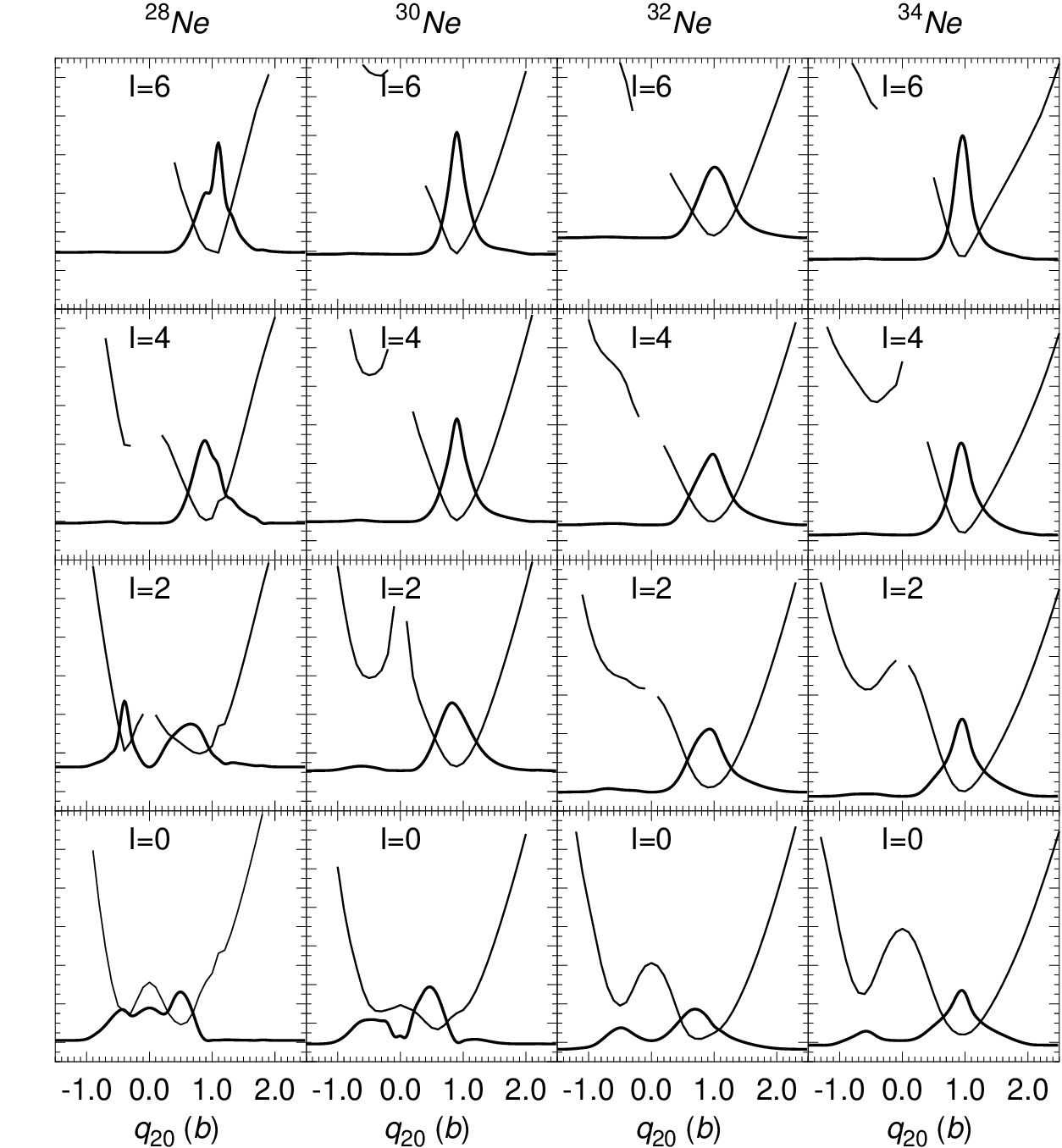}} \par}

\caption{Collective wave functions squared for the ground states
(\protect\( \sigma =1\protect \)) and the spin values \protect\( I^{\pi }=0^{+},2^{+},4^{+}\protect \)and
\protect\( 6^{+}\protect \) of the nuclei \protect\( ^{20-34}\protect \)Ne.
The corresponding projected energy curve is also plotted for each
spin value. The \protect\( y-\protect \)axis scales are in energy
units and always span an energy interval of \protect\( 13\protect \)
MeV (minor ticks are \protect\( 0.5\protect \) MeV apart). The collective
wave functions have also been plotted against the energy scale after
proper scaling and shifting, that is, the quantity \protect\( E^{I}_{\sigma }+15\times \mid g_{\sigma }^{I}(q_{20})\mid ^{2}\protect \)
is the one actually plotted. With this choice of the scales we can
read from the figure the energy gain due to quadrupole fluctuations
by considering the position of the wave functions' tail relative to
the projected curve.}\label{Fig_8_art}
\end{figure*}

In Fig. \ref{Fig:BandDiagram} we have plotted the AMP-GCM energies \( E_{\sigma }^{I} \)
obtained by solving the Hill-Wheeler equation  and placed them along the 
\( q_{20} \)-axis
according to their ``average intrinsic quadrupole moment"  
defined as 
\begin{equation}
\overline{q}_{20}^{I,\sigma }=
\int dq_{20}\mid g^{I}_{\sigma }(q_{20})\mid ^{2}q_{20}
\end{equation}

In this figure we have also plotted the \( I^{\pi }=0^{+} \) AMPPES to guide
the eye. The first noticeable fact is that the correlated $0^+$ ground states 
have a lower energy than the absolute minimum of the corresponding AMPPES. This
energy gain due to the quadrupole correlations increases the binding energies
and can be relevant for a proper description of the two neutron separation
energies. The second prominent feature is that configuration mixing decreases 
the deformation of the $0^+_1$ ground states with respect to the minimum of the
AMPPES. The ground states of the nuclei $^{24}$Ne and $^{26}$Ne  become
spherical whereas the ones of $^{28}$Ne and $^{30}$Ne become weakly deformed.
The other nuclei remain well deformed in their ground states and develop a
rather well defined rotational band up to the maximum spin considered for
$\sigma=1$. In addition, a well defined rotational band is obtained for $I \ge
2$ in $^{30}$Ne. On the other hand, the excited states ($\sigma=2$) only show a
rotational band pattern for those nuclei well  deformed in their ground state.
Another interesting point concerns the energy gain due to configuration mixing
of the ground state: it is 0.6 MeV for $^{20}$Ne, increases up to 0.9 MeV  for
$^{24}$Ne and $^{26}$Ne and then monotonically decreases up to 0.55 MeV in
$^{34}$Ne. The energy gain is correlated with the shape of the AMPPES for I=0,
namely, if the AMPPES shows a well defined minimum the energy gain is smaller
than in the case where the minimum is broader.

To gain a deeper insight into the intrinsic structure of the AMP-GCM sates the
ground band (\( \sigma =1 \)) collective wave functions squared \( \mid
g^{I}_{\sigma }(q_{20})\mid ^{2} \) are plotted in Fig. \ref{Fig_8_art} along
with the corresponding AMPPES.  For \( ^{20} \)Ne and \( ^{22} \)Ne the ground
state collective wave functions are well inside the prolate wells indicating
that both systems are dominated by prolate deformations in the considered spin
range. The average deformation values are 0.39b, 0.58b, 0.59b and 0.58b for the
\( 0_{1}^{+} \), \( 2_{1}^{+} \), \( 4_{1}^{+} \) and \( 6_{1}^{+} \) states in
\( ^{20} \)Ne, while the corresponding values in \( ^{22} \)Ne are 0.39b,
0.56b, 0.61b and 0.63b. In the nucleus \( ^{24} \)Ne, the configuration mixing
calculation provides a spherical \( 0_{1}^{+} \) instead of the oblate absolute
minimum found in its AMPPES. An almost spherical \( 2_{1}^{+} \) state (\(
\overline{q}_{20}^{I=2,\sigma =1}=-0.08 \)b) is also found in this nucleus. On
the other hand, a band crossing takes place for the state \( 4_{1}^{+} \) and
the collective wave function becomes prolate deformed with \(
\overline{q}_{20}^{I=4,\sigma =1}=0.39 \)b and \( \overline{q}_{20}^{I=6,\sigma
=1}=0.79 \)b. Similar results have recently been found \cite{Rayner.2002} in
the \( N=14 \) system \( ^{26} \)Mg. The experimental spectroscopic quadrupole
moment of the lowest \( 2^{+} \) state in \( ^{26} \)Mg is -13.5 \(
e\textrm{fm}^{2} \) \cite{Endt.98} indicating that this is a prolate state.
However, the low excitation energy of the \( 0_{2}^{+} \) in \( ^{26} \)Mg
(3.588 MeV) compared with the same quantity in \( ^{24} \)Mg (6.432 MeV)
clearly indicates strong shape coexistence between oblate and prolate
solutions. Unlike \( ^{26} \)Mg, there is not experimental information
concerning the spectroscopic quadrupole moment of the lowest \( 2^{+} \) state
in \( ^{24} \)Ne but again the experimental excitation energies of the \(
0_{2}^{+} \) in \( ^{24} \)Ne and \( ^{26} \)Ne (4.764 MeV and 3.691 MeV,
respectively) are significantly lower than   the 6.235 MeV measured in \( ^{22}
\)Ne \cite{Endt.98,Endt.90} pointing towards a strong shape coexistence in the
ground state of the former nuclei. Our results predict a strong shape
coexistence for the \( 0_{1}^{+} \) and \( 2_{1}^{+} \) states in \( ^{24} \)Ne
as well as for the \( 0_{1}^{+} \) states in \( ^{26} \)Ne and \( ^{28} \)Ne
that manifest itself in the \( 0^{+}_{2} \) excitation energies of 3.509 and
3.005 MeV for \( ^{24,26} \)Ne, respectively. 

The \( 0_{1}^{+} \) wave functions in both \( ^{26} \)Ne and \( ^{28} \)Ne show
a great admixture of prolate and oblate configurations that leads to spherical
ground states on the average. On the other hand, for \( I^{\pi
}\geq 2^{+} \) the collective wave functions in these two nuclei become prolate
deformed with deformation values of 0.34b, 0.59b and 0.94b for the states \(
2_{1}^{+} \), \( 4_{1}^{+} \) and \( 6_{1}^{+} \) in \( ^{26} \)Ne while the
corresponding values in \( ^{28} \)Ne are 0.26b, 0.93b and 1.04b. Our results
do not fully support  the suggestion of \cite{Ozawa:2000} concerning the
magicity of N=16 in neutron rich light nuclei. It is true that $^{26}$Ne has a
spherical ground state, but the deformation of the $2^+_1$ is too strong as to
be considered a vibrational state. Similar results are also found in Ref. 
\cite{Rayner.2002} for the N=16 nucleus $^{28}$Mg.

The \( 0_{1}^{+} \) wave function in \( ^{30} \)Ne also shows a
significant admixture of the oblate and prolate configurations
and, as a consequence, the deformation in the ground state is
reduced to 0.16 b that represents one third of the value
corresponding to the absolute minimum in the \( I^{\pi }=0^{+} \)
AMPPES. This clearly shows that, as in \( ^{32} \)Mg
\cite{Heenen.99,Rayner.2000.a,Rayner.2002}, the deformation
effects in \( ^{30} \)Ne are the result of a subtle balance
between the zero point corrections associated with the
restoration of the rotational symmetry and the fluctuations in
the collective parameters (in our case the axially symmetric
quadrupole moment). From the comparison with the value of \(
\overline{q}_{20}^{I=0,\sigma =1}=0.43 \) b already found
\cite{Rayner.2000.a,Rayner.2002} in \( ^{32} \)Mg we conclude
that dynamical deformation effects are strongly suppressed in \(
^{30} \)Ne as could have been forecasted from the different
AMPPES topology we have already seen in
Fig.\ref{Fig:30Neand32Mg}. On the other hand, the \( 2_{1}^{+}
\), \( 4_{1}^{+} \) and \( 6_{1}^{+} \) wave functions in \(
^{30} \)Ne are inside the prolate wells and the average
deformations of 0.83b, 0.93b and 0.94b are very close to the ones found
in \( ^{32} \)Mg.

\begin{table}
\caption{Ground state spectroscopic quadrupole moments \protect\(
Q^{spec}(I,\sigma =1)\protect \) in \protect\( e\textrm{fm}^{2}\protect \) for
\protect\( I^{\pi }=2^{+}\protect \), \protect\( 4^{+}\protect \) and
\protect\( 6^{+}\protect \) in the nuclei \protect\( ^{20-34}\protect \)Ne. 
Experimental data \cite{Endt.98,Raghavan.89} are shown in boldface whereas
Shell Model results \cite{Caurier.98,Caurier.01} are shown in brackets. For
details see the main text. \label{Table:Qspec}}
\begin{center}
\begin{tabular}{|c|c|c|c|c|}\hline 
I & $^{20}$Ne & $^{22}$Ne & $^{24}$Ne & $^{26}$Ne \\ \hline \hline 
2 & -16.75    & -14.64    &   2.02    &  -7.88    \\		   
  & {\bf -23} & {\bf -17} &(-2.77)    &(-10.56)   \\ \hline	   
4 & -21.52    & -21.03    & -11.62    & -16.98    \\ \hline	   
6 & -23.29    & -23.01    & -27.01    & -28.78    \\ \hline \hline

I  & $^{28}$Ne & $^{30}$Ne & $^{32}$Ne & $^{34}$Ne \\ \hline \hline 
2  & -10.05	& -15.80    & -17.49	& -15.81    \\  	     
   &(-17.8)	&(-16.40)   &(-20.00)	&(-16.56)   \\ \hline	     
4  & -26.11	& -22.21    & -22.25	& -21.38    \\ \hline	     
6  & -29.83	& -24.82    & -26.23	& -23.52    \\ \hline	     
\end{tabular}
\end{center}
\end{table}
In both \( ^{32} \)Ne and \( ^{34} \)Ne, the ground state collective
wave functions become prolate. The dynamical deformation values for
the \( 0_{1}^{+} \), \( 2_{1}^{+} \), \( 4_{1}^{+} \) and \( 6_{1}^{+} \)
in \( ^{32} \)Ne are 0.42b, 0.90b, 0.98b, 1.05b while the corresponding
values in \( ^{32} \)Ne are 0.69b, 0.94b, 0.99b and 1.0b. All these
values show the stability of deformation effects in Neon isotopes
as we move towards the drip-line.

Here we will make a few comments on the AMP-GCM results in the nuclei \( ^{32}
\)Ne and \( ^{34} \)Ne when the basis is increased from \( N_{Shell}=11 \) to
\( N_{Shell}=12 \). The effect on the projected energy landscapes can be seen
in the corresponding panels of Fig. \ref{Fig:AMPPES} where \( N_{Shell}=12 \)
curves are plotted as dashed lines. The results with \( N_{Shell}=12 \) and \(
N_{Shell}=11 \) only show very small differences at large absolute values of \(
q_{20} \). On the other hand, in the region of physical significance (\(
-.7\textrm{b}\leq q_{20}\leq 1.2\textrm{b} \)) the results are practically
indistinguishable. Since this range of \( q_{20} \) values is the one where the
collective dynamics is concentrated (i.e, the tails of the collective wave
functions go to zero outside this range) we do not expect big changes in the
excitation energies. For the $^{32}$Ne nucleus, it turns out that the 
excitation energies of all the members of
the ground state rotational band obtained in the \( N_{Shell}=12 \) calculation
are around 7 keV
higher than the ones in the  \( N_{Shell}=11 \) case, i.e., only the \(
0^{+}_{1} \) state has been pushed down. For $^{34}$Ne the average shift is 6
keV. As a consequence, the transition gamma ray energies remain unaltered by the
increase of the basis size. On the other hand, the excitation energies (with
respect to the true ground state) of the members of the excited rotational band
decrease on the average 40 keV in \( ^{32} \)Ne and 32 keV in \( ^{34} \)Ne and
therefore, as in the previous case, the intra-band gamma ray energies remain
the same. From these results we conclude that our calculations are well
converged in terms of the basis size.

\begin{figure}
{\centering \resizebox*{0.5\textwidth}{!}{\includegraphics{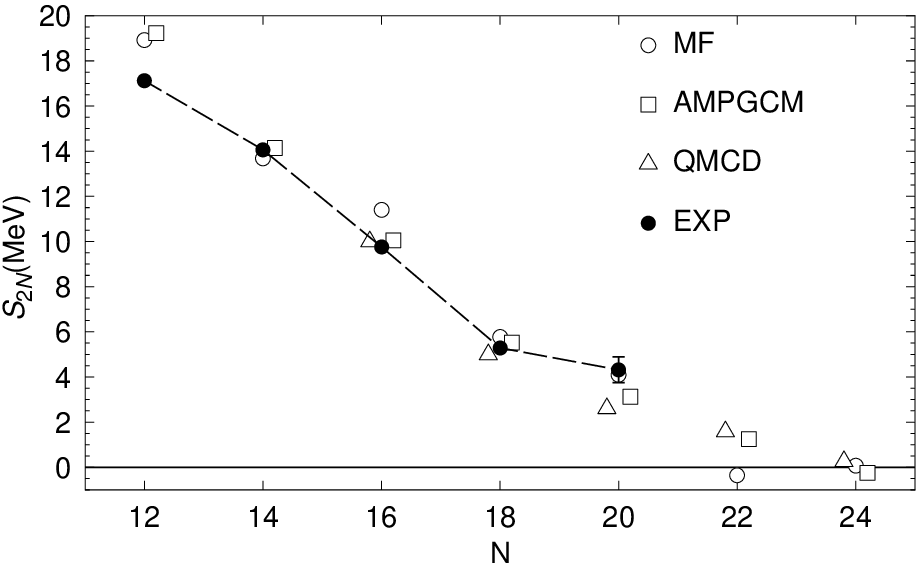}} \par}
\caption{The two neutron separation energies obtained with the
different approximations considered in this paper are compared to
experimental values taken from Ref. \cite{AUDI,Sarazin.00}. and the QMCD results
of Ref. \cite{Utsumo.99,OTSUKA_REVIEW}}
\label{Fig:S2N}
\end{figure}

\begin{figure*}
{\centering \resizebox*{0.9\textwidth}{!}{\includegraphics{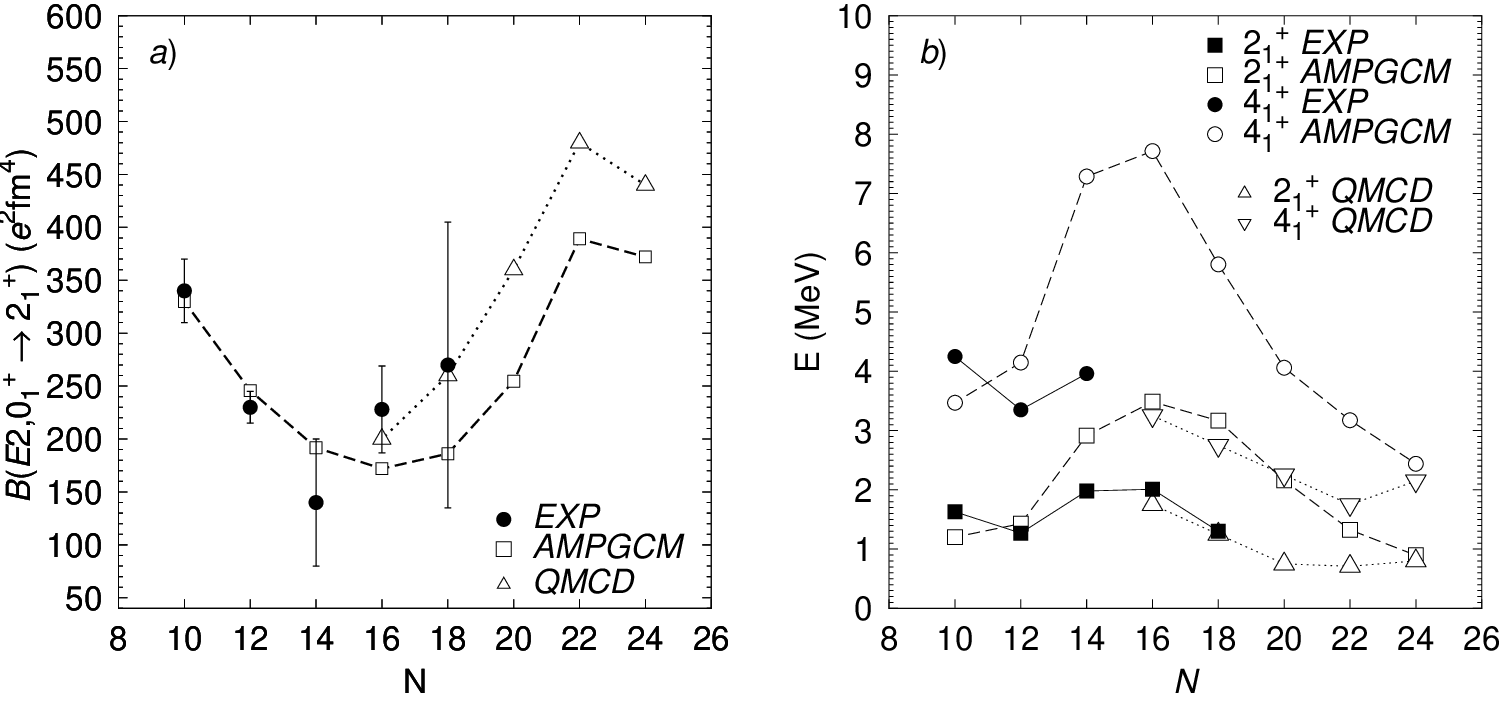}} \par}

\caption{On the left panel the experimental values \cite{Endt.90,Endt.98,Reed.99,Prity.99,Azaiez.99,Guillemaud.02}
of the \protect\( B(E2,0^{+}_{1}\rightarrow 2^{+}_{1})\protect \)
transition probabilities are compared to our predictions and other
theoretical results \cite{Utsumo.99,OTSUKA_REVIEW}. On the right
panel, the same comparison is made for the \protect\( 2^{+}_{1}\protect \)
and \protect\( 4^{+}_{1}\protect \) excitation energies.
\label{Fig:EnergiesandBE2}}
\end{figure*}

Coming back to the discussion of our results, in Table \ref{Table:Qspec} we
present our results for the ground band spectroscopic quadrupole moments. A
very good agreement is observed between the calculated spectroscopic values of
the \( 2_{1}^{+} \) states in \( ^{20} \)Ne and \( ^{22} \)Ne and the
experimental values (shown in boldface) for these nuclei which are -23 \(
e\textrm{fm}^{2} \) \cite{Raghavan.89} and -17 \( e\textrm{fm}^{2} \)
\cite{Endt.98}, respectively. In addition,  our results are consistent  with
the Shell Model predictions (shown in brackets) for the spectroscopic
quadrupole moments of the \( 2_{1}^{+} \) states \cite{Caurier.98,Caurier.01}. 
Obviously, the discrepancies are larger for the shape coexistent nuclei \(
^{24-28} \)Ne,  It is also worth mentioning here that the value obtained for
the spectroscopic quadrupole moment of the \( 2_{1}^{+} \) state in \( ^{30}
\)Ne (-15.80 \( e\textrm{fm}^{2} \)) is only slightly smaller than the similar
quantity in \( ^{32} \)Mg (-19.50 \( e\textrm{fm}^{2} \)).

In Fig. \ref{Fig:S2N} we compare the results of the AMP-GCM two neutron
separation energies \( S_{2N}=E_{0_{1}^{+}}(N-2)-E_{0_{1}^{+}}(N) \) with the
corresponding mean field results (see subsection \ref{MF_RESULTS}) and also
with the available experimental values taken from Ref. \cite{AUDI,Sarazin.00}.
The AMP-GCM improves the $S_{2N}$ of $^{26}$Ne as compared to the mean field
results and also makes bound (unlike the mean field prediction) the nucleus
$^{32}$Ne in good agreement with the experimental results
\cite{GANIL_1,GANIL_2,SAKU_1}. On the other hand, the nucleus $^{34}$Ne becomes
unstable against two neutron emission in the AMP-GCM approach. Our results are
very similar to  the ones predicted in the framework of the QMCD
\cite{Utsumo.99,OTSUKA_REVIEW} specially around \( N=20 \) but we differ in the
prediction concerning  $^{34}$Ne. Finally, it is worth mentioning that the
AMP-GCM binding energy is the sum of the mean field binding energy of the
intrinsic state plus the energy gain due to the restoration of the rotational
symmetry plus the energy gain due to the configuration mixing. Therefore, the
differences in \( S_{2N} \) obtained in the AMP-GCM and the mean field are due
to the last two contributions. The analysis of those contributions shows that
the rotational energy correction is the main responsible for the differences
observed in \( S_{2N} \) and therefore is the ingredient needed to make
$^{32}$Ne stable.

In Fig. \ref{Fig:EnergiesandBE2} the excitation energies of the \( 2_{1}^{+}
\) and \( 4_{1}^{+} \) states and the \( B(E2,0_{1}^{+}\rightarrow 2_{1}^{+})
\) transition probabilities obtained in the AMP-GCM are compared with the
available experimental values and also with the predictions of the QMCD
\cite{Utsumo.99,OTSUKA_REVIEW}. Concerning the \( B(E2,0_{1}^{+}\rightarrow
2_{1}^{+}) \) transition probabilities we clearly see, from panel a), that
the agreement with the available experimental data
\cite{Endt.90,Endt.98,Reed.99,Prity.99,Azaiez.99,Guillemaud.02} is rather
satisfactory and in most cases (with the exception of \( ^{26} \)Ne where our
prediction appears a little bit low) our results stay within the experimental
error bars. On the other hand, our results are also consistent with the
predictions of the QMCD \cite{Utsumo.99,OTSUKA_REVIEW}. Our results although
not as good as the QMCD ones, are very satisfactory considering that the
parameters of the Gogny force have not been fitted to the region and/or the
physics of quadrupole collectivity and also that no effective charges have
been used in our calculations of the transition probabilities.

Our results for the \( 2^{+}_{1} \) excitation energies agree well with the
experiment in \( ^{20} \)Ne and \( ^{22} \)Ne \cite{Endt.90,Endt.98} and also
with the theoretical result of Otsuka in \( ^{34} \)Ne. For the
other isotopes our \( 2^{+} \)excitation energies are always higher than
experiment and the QMCD predictions. The same happens for the \( 4^{+} \)
excitation energies. The disagreement between our results and the experiment
(or the QMCD predictions) can be attributed to the fact that the nuclei
involved are nice examples of shape coexistence. In the presence of shape
coexistence the quadrupole degree of freedom probably is not enough to
describe accurately the observables and other degrees of freedom like
triaxiality or pairing fluctuations should be included. However, and based on
the nice agreement between our electromagnetic transition probabilities and
the experimental ones, we can conclude that the quadrupole degree of freedom
is the main ingredient in the physical picture of the neutron rich Neon
isotopes.

\section{Conclusions}

We have performed angular momentum projected Generator
Coordinate Method calculations with the Gogny interaction D1S and the mass
quadrupole moment as generating coordinate in order to describe quadrupole
collectivity in the even-even nuclei \( ^{20-34} \)Ne. The lighter isotopes
$^{20}$Ne and $^{22}$Ne as well as the heavier ones $^{32}$Ne and $^{34}$Ne
are well deformed in their ground states whereas the other isotopes are
either spherical ($^{24}$Ne and $^{26}$Ne) or slightly deformed like $^{28}$Ne
and  $^{30}$Ne. The later isotope is less deformed in its ground state than
its isotone \( ^{32} \)Mg. The two neutron separation energies  compare
well with experimental data and show that \( ^{32} \)Ne is bound. Moreover,
the \( B(E2) \) transition probabilities from the ground state to the \( 2^{+}
\) state are well reproduced in all the cases. Only the $2^+$ excitation
energies come too high as compared to the experiment for the shape coexistent
isotopes.

\begin{acknowledgement}

This work has been supported in part by the DGI, Ministerio de Ciencia y
Tecnología (Spain) under project BFM2001-0184.
\end{acknowledgement}

\end{document}